\documentclass[letterpaper,conference,onecolumn]{IEEEtran}
\IEEEoverridecommandlockouts
\usepackage{siunitx}
\usepackage{ifthen}
\usepackage{amsmath,amssymb,amsfonts}
\usepackage{algorithmic}
\usepackage{graphicx}
\usepackage{textcomp}
\usepackage{colortbl}
\usepackage{ragged2e}
\usepackage{tabularx}
\usepackage{multirow}
\usepackage{float}
\usepackage{hyperref}
\usepackage{comment}
\usepackage{tikz}
\usepackage{circuitikz}
\usepackage{dirtytalk}
\usepackage[nolist]{acronym}
\usepackage{subcaption}
\usepackage[font=small]{caption}
\usepackage{makecell, tabularx}
\usepackage{xcolor}
\usepackage{tabularray}
\usepackage[boxed,commentsnumbered, linesnumbered,ruled,vlined]{algorithm2e}

\usepackage{listings}

\usepackage{pgfplots}
\usepackage{tikz}
\usepackage{circuitikz}
\usetikzlibrary{calc,trees,positioning,arrows,chains,shapes.geometric,%
	decorations.pathreplacing,decorations.pathmorphing,shapes,%
	matrix,shapes.symbols}
\graphicspath{{./figures/}}
\tikzstyle{startstop} = [rectangle, rounded corners, minimum width=3cm, minimum height=1cm,text centered, draw=black, fill=red!30]
\tikzstyle{io} = [trapezium, trapezium left angle=70, trapezium right angle=110, minimum width=3cm, minimum height=1cm, text centered, draw=black, fill=blue!30]
\tikzstyle{process} = [rectangle, minimum width=3cm, minimum height=1cm, text centered, text width=4cm, draw=black, fill=orange!30]
\tikzstyle{decision} = [diamond, minimum width=3cm, minimum height=1cm, text centered, draw=black, fill=green!30]
\tikzstyle{arrow} = [thick,->,>=stealth]

\usepackage[ruled,linesnumbered]{algorithm2e}
\SetAlCapNameFnt{\small}
\SetAlCapFnt{\small}
\SetAlgoNlRelativeSize{0}
\RestyleAlgo{ruled}
\SetKwComment{Comment}{/* }{ */}

\usepackage[style=ieee, backend=biber, natbib=true]{biblatex}
\newboolean{blindreview}
\setboolean{blindreview}{false}

\begin{document}
	

	\begin{acronym}[placeholder]		
		\acro{SoCs}{System-on-Chips}
		\acro{ECCs}{Error Correction Code}
		\acro{ECC}{Error Correction Code}
		\acro{SECDED}{Single-Error Correction, Double-Error Detection}
		\acro{IPs}{Intellectual Properties}
		\acro{HDLs}{Hardware Description Languages}
		\acro{HVLs}{Hardware Verification Language}
		\acro{CRV}{Constrained Random Verification}
		\acro{BFMs}{Bus Functional Models}
		\acro{SoC}{System-on-Chip}
		\acro{FV}{Formal Verification}
		\acro{SAR}{Successive Approximation Register}
		\acro{acronym}{full name}
		\acro{GPI}{General Purpose Interface}
		\acro{MDV}{Metric Driven Verification}
		\acro{HDL}{Hardware Description Language}
		\acro{HVL}{Hardware Verification Language}
		\acro{ADC}[ADC]{Analog-to-Digital Converter}
		\acro{I2C}{Inter-Integrated Circuit}
		\acro{ALU}{Arithmetic Logic Unit}
		\acro{ASIC}{Application-Specific Integrated Circuit}
		\acro{AI}{Artificial Intelligence}
		\acro{IP}{Intellectual Property}
		\acro{Cocotb}{Coroutine based cosimulation testbench}
		\acro{TLM}{Transaction Level Modeling}
		\acro{UVM}{Universal Verification Methodology}
		\acro{DUT}{Design Under Test}
		\acro{BFM}{Bus Functional Model}
		\acro{EDA}{Electronic Design Automation}
		\acro{RAL}{Register Abstraction Layer}
		\acro{VPI}{Verilog Procedural Interface}
		\acro{VHPI}{VHDL Procedural Interface}
		\acro{FPGA}{Field Programmable Gate Array}
		\acro{PyUVM}{Python Universal Verification Methodology}
	\end{acronym}


\lstset{
	language=Verilog,           
	basicstyle=\footnotesize,   
	numbers=left,               
	frame=lines,                
	captionpos=b,               
	breaklines=true,            
	tabsize=2,                  
	xleftmargin=0pt,
	framexleftmargin=0pt
}

\lstdefinestyle{Python}{
	language        = Python,
	basicstyle      = \ttfamily,
	keywordstyle    = \color{blue},
	keywordstyle    = [2] \color{teal}, 
	stringstyle     = \color{green},
	commentstyle    = \color{red}\ttfamily
}

\lstset{
	frame       = single,
	numbers     = left,
	showspaces  = false,
	showstringspaces    = false,
	captionpos  = t,
	caption     = \lstname
}



	\title{Towards Efficient Design Verification – Constrained Random Verification using PyUVM\\}

	\ifthenelse{\boolean{blindreview}}{}{
		\author{\IEEEauthorblockN{Deepak Narayan Gadde}
				\IEEEauthorblockA{Infineon Technologies \\
				Dresden, Germany \\
				Deepak.Gadde@infineon.com}
				\and
				\IEEEauthorblockN{Suruchi Kumari}
				\IEEEauthorblockA{Infineon Technologies \\
					Dresden, Germany \\
					Suruchi.Kumari@infineon.com}
				\and
				\IEEEauthorblockN{Aman Kumar}
				\IEEEauthorblockA{Infineon Technologies \\
				Dresden, Germany \\
				Aman.Kumar@infineon.com}
				}
	}
	

	\maketitle
	
	

	\begin{abstract} 
		\textbf{\emph{Abstract}\!
			\textemdash Python, as a multi-paradigm language known for its ease of integration with other languages, has gained significant attention among verification engineers recently. A Python-based verification environment capitalizes on open-source frameworks such as \textit{PyUVM} providing Python-based UVM 1.2 implementation and \textit{PyVSC} facilitating constrained randomization and functional coverage. These libraries play a pivotal role in expediting test development and hold promise for reducing setup costs. The goal of this paper is to evaluate the effectiveness of PyUVM verification testbenches across various design IPs, aiming for a comprehensive comparison of their features and performance metrics with the established SystemVerilog-UVM methodology.
			}
		
	\end{abstract}

	\section{Introduction}\label{sec:intro}
	With the continuous increase in complexity of \ac{SoC} designs, verification is becoming ever more challenging. As a result, the time required for verification experiences a significant upsurge. Additionally, there is a subsequent need to be more productive and efficient with limited manpower. Industry-utilized methodologies like \ac{CRV}, \ac{FV}, and \ac{MDV} use SystemVerilog as a language construct. It provides numerous features like object-oriented programming and functional coverage, but learning the language has a steep curve especially for the freshers requiring good understanding of designs.

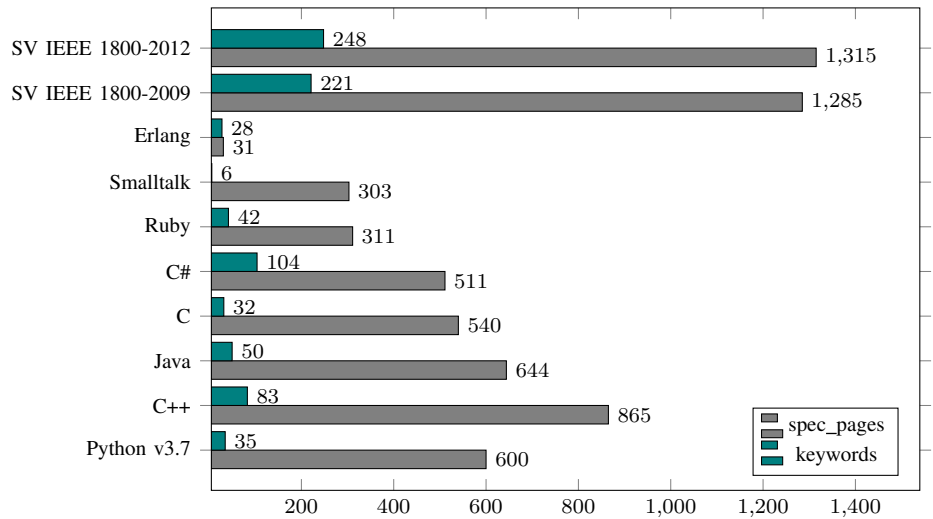
\begin{figure}[H]
	\centering
	\begin{tikzpicture}
	\begin{axis}[
		xbar=0pt,
		xmin=5,
		width=11cm,
		height=8cm,
		xmax=1400,
		bar width=7pt,
		enlarge x limits={value=0.1,upper},
		legend pos={south east,font=\footnotesize},
		symbolic y coords={Python v3.7,C++, Java, C,C\#,Ruby, Smalltalk, Erlang, SV IEEE 1800-2009, SV IEEE 1800-2012},
		ytick=data,
		y tick label style={anchor=east,font=\footnotesize},
		nodes near coords,
		nodes near coords align={horizontal}
		]
		\addplot [fill=gray,font=\footnotesize] coordinates {(600,Python v3.7) (865,C++) (644,Java) (540,C) (511,C\#) (311,Ruby) (303,Smalltalk) (31,Erlang) (1285,SV IEEE 1800-2009) (1315,SV IEEE 1800-2012)};
		\addplot [fill=teal,font=\footnotesize] coordinates {(35,Python v3.7) (83,C++) (50,Java) (32,C) (104,C\#) (42,Ruby) (6,Smalltalk) (28,Erlang) (221,SV IEEE 1800-2009) (248,SV IEEE 1800-2012)};
		\legend{spec\_pages,keywords}
	\end{axis}
\end{tikzpicture}
	\caption[Language complexity with respect to number of specification pages and keywords]{Language complexity with respect to number of specification pages and keywords \cite{lang_complex}}
	\label{lang_complex}
\end{figure}

Fig. \ref{lang_complex} shows SystemVerilog is the most complicated language in comparison with other programming languages, as it has 1315 specification pages and 248 keywords as per IEEE 1800-2012, with variations in the level of tool support across different \ac{EDA} vendors \cite{lang_complex}. In summary, \ac{UVM}, which utilizes SystemVerilog, gets more complex in addition to its standard. On the other hand, Python is simple, less verbose, and easily integrable to Numpy, Pandas, and other open-source libraries. A recent study conducted by the Wilson research group \cite{VerStudy_22} has shown that \SI{23}{\percent} of all \ac{ASIC} projects used Python for various project specific tasks.  

Like SystemVerilog-\ac{UVM}, PyUVM implements \ac{UVM} 1.2 IEEE specification in Python using its high level language features, and is built upon Cocotb to interface with the simulators. The main objective of this work is to assess the effectiveness of Python testbench development using PyUVM and PyVSC libraries for the design \ac{IPs}, and to compare it with the existing state-of-the-art methodologies, such as \ac{UVM} and \ac{CRV}. The following are key highlights of the paper:

\begin{itemize}
	\item Comparison of Systemverilog-\ac{UVM} with PyUVM for testbench development of various design \ac{IPs}
	\item Feasibility of PyVSC library for enhanced coverage declaration and \ac{CRV}
	\item Comparison of performance metrics in terms of simulation run-time
	\item Empirical observations made during the development of Python testbenches 
\end{itemize}

The structure of the paper is as follows: Section \ref{sec:background} summarizes various prior works related to python based testbenches. Section \ref{sec:designs} details the design \ac{IPs} used for the testbench development in both Systemverilog-\ac{UVM} and PyUVM. Section \ref{sec:implementation} explains the implementation of our approach in creating PyUVM testbenches with an example. Section \ref{sec:results} discusses the results and empirical observations made during our research. At the end, Section \ref{sec:conclusion} concludes our current work with future scope.

	\section{Related Work}\label{sec:background}
	Prior work \cite{merldsu} has created the PyUVM framework for a RISC-V Single Cycle Core and has encouraged the features supported by the methodology. The author in \cite{ian_quinn} discusses how the verification environment created in Python helps to reuse the test sequences across testbenches. In \cite{verifog}, the author proposed a flow using the \textit{Verifog} tool to catch bugs at the earliest stages of the design phase without developing testbenches. \textit{uvm-python} is a port of SystemVerilog-\ac{UVM} 1.2 to Python and Cocotb tested with Icarus Verilog and Verilator \cite{Poikela}. Additionally, research \cite{suruchi} has compared Python with SystemVerilog to show its significance in design verification. The authors provide a detailed comparison between both \acp{HVL} in terms of design hierarchy, coverage constructs, and their performance during simulation. The feature comparison from the paper is given below in Table \ref{sv_python}.

\begin{table}[H]
	\centering
	\caption{Comparison between SystemVerilog and Python \cite{suruchi}}
	\label{sv_python}
	\setlength{\tabcolsep}{5pt}
	\resizebox{\textwidth}{!}{%
		\begin{tabular}{|l|l|l|l|} 
			\hline
			\textbf{Feature}                                                                                      & \textbf{SystemVerilog}                                           & \textbf{Python}                                                          & \textbf{Remarks}                                                                                                                                                                                             \\ 
			\hline
			\begin{tabular}[c]{@{}l@{}}\textbf{Declaration of }\\\textbf{data types}\end{tabular}                 & Static                                                     & \begin{tabular}[c]{@{}l@{}}Dynamic
			\end{tabular}                  & \begin{tabular}[c]{@{}l@{}}Python allows undeclared variables and perform \\any operation on them. Additionally, it has advanced data structures \\ e.g., tuple and dictionary, unlike SystemVerilog.\end{tabular}                \\ 
			\hline
			\begin{tabular}[c]{@{}l@{}}\textbf{Supported types }\\\textbf{of logic}\end{tabular}                  & \textit{0, 1, X, Z}                                              & \textit{X, Z, U, W}                                                      & Python-\ac{Cocotb} needs BinaryValue object for these logics                                                                                                                                                      \\ 
			\hline
			\begin{tabular}[c]{@{}l@{}}\textbf{Parameterization and }\\\textbf{size of the variable}\end{tabular} & Required                                                         & Not required                                                             & \begin{tabular}[c]{@{}l@{}}If size is not declared in SystemVerilog, data may be \\lost after an assignment to a different size than specified\end{tabular}                                                  \\ 
			\hline
			\textbf{Styles of control flow}                                                                       & \textit{begin, end}                                       & Proper indentation                                                       & \textit{elif }in Python replaces case in SystemVerilog/Verilog                                                                                                                                               \\ 
			\hline
			\textbf{Functions}                                                                                    & Not objects                                                      & Callable objects                                                         & \begin{tabular}[c]{@{}l@{}}Function in SystemVerilog are not objects and cannot be \\stored or passed directly as arguments\end{tabular}                                                                     \\ 
			\hline
			\textbf{Exceptions}                                                                                   & Not supported                                                    & Supported                                                                & In Python, exceptions are caught with \textit{try/except/finally} blocks                                                                                                                                     \\ 
			\hline
			\textbf{Libraries}                                                                                    & -                                                                & Large                                                                    & \textcolor[rgb]{0.114,0.114,0.114}{Create reference model for any complex design easily}                                                                                                                     \\ 
			\hline
			\textbf{Interpreted}                                                                                  & No                                                               & Yes                                                                      & \begin{tabular}[c]{@{}l@{}}\textcolor[rgb]{0.114,0.114,0.114}{It allows to restart the simulator without recompiling and }\\\textcolor[rgb]{0.114,0.114,0.114}{edit tests while it is running}\end{tabular}  \\ 
			\hline
			\textbf{Design Hierarchy}                                                                             & \begin{tabular}[c]{@{}l@{}}Includes top \\testbench\end{tabular} & \begin{tabular}[c]{@{}l@{}}Does not include \\top testbench\end{tabular} & \begin{tabular}[c]{@{}l@{}}It limits debugging capabilities, since tracing back signals \\in the testbench is not possible\end{tabular}                                                                      \\
			\hline
		\end{tabular}%
	}
\end{table}

The related works elaborated on above mostly discuss PyUVM implementation but none of them examines how it is different from commonly used methodology i.e., SystemVerilog-UVM in functional verification. In this work, we try to address this by comparing various features of PyUVM with SystemVerilog-\ac{UVM}, and their performances are analyzed in terms of simulation run-time.
	
	\section{Designs}\label{sec:designs}
	We have carefully chosen three distinct designs in our research for verification implementation considering several important aspects i.e., \ac{ALU}, \ac{ADC}, and \ac{ECC}. These may aid in our decision about compatibility of PyUVM testbenches in comparison with SystemVerilog-\ac{UVM} methodology. A short explanation for the design \ac{IPs} is given below.

\subsection{ALU}
\begin{figure}[H]
	\centering
	\includegraphics[scale=0.65]{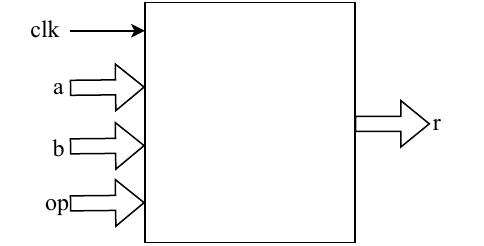}
	\caption{ALU}
	\label{design_alu}
\end{figure}

The 32-bit \ac{ALU} is a hardware unit that is designed using Verilog and can perform various computations on 32-bit input data. These include two arithmetic operations: addition and subtraction; and six logical operations, including NOT, AND, OR, XOR, NAND, and NOR. This \acs{IP} has 4 inputs, two of which are data buses \textit{a} and \textit{b}, plus a control bus \textit{op}, and a clock signal. The output of the design module would be a data bus \textit{r}. Fig. \ref{design_alu} displays the block diagram of the 32-bit \ac{ALU}. Due to its computational capabilities, it is commonly used in modern processors.

\subsection{ADC}

\begin{figure}[H]
	\centering
	\includegraphics[scale=0.65]{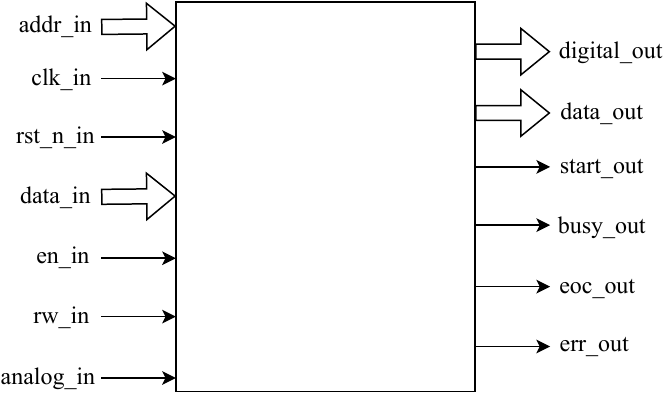}
	\caption{ADC}
	\label{design_adc}
\end{figure}

The \ac{ADC} module is primarily meant to convert analog input values to the corresponding digital data with 16 bit resolution. To improve the resolution and reduce the noise of the conversion, an oversampling feature is also added to the \ac{ADC}. Control and Status Registers are available to configure the oversampling factor of 1, 2, 4, or 8, with 1 indicating no oversampling. The conversion of \textit{analog\_in} can be triggered via another register bit and the result sent out on the \textit{digital\_out} bus as shown in Fig. \ref{design_adc}.

\begin{table}[H]
	\centering
	\setlength{\extrarowheight}{3pt}
	\setlength{\tabcolsep}{5pt}
	\caption{Description of Register Model used in ADC design IP}
	\label{adc_reg}
	\begin{tabular}{|l|l|l|l|l|l|}
		\hline
		\textbf{Register}        & \textbf{Address}     & \textbf{Width}     & \textbf{Access} & \textbf{Reset Value} & \textbf{Details}              \\ \hline
		Dummy                    & 'h0                  & 8                  & RO              & 'h0                  & Dummy register - not used     \\ \hline
		\multirow{2}{*}{Config}  & \multirow{2}{*}{'h1} & \multirow{2}{*}{8} & {[}7:2{]} RO    & \multirow{2}{*}{'h0} & Reserved - not used           \\ \cline{4-4} \cline{6-6} 
		&                      &                    & {[}1:0{]} RW    &                      & Oversampling factor (1,2,4,8) \\ \hline
		\multirow{2}{*}{Trigger} & \multirow{2}{*}{'h2} & \multirow{2}{*}{8} & {[}7:1{]} RO    & \multirow{2}{*}{'h0} & Reserved - not used           \\ \cline{4-4} \cline{6-6} 
		&                      &                    & {[}0{]} RW      &                      & Start ADC conversion          \\ \hline
	\end{tabular}
	
\end{table}

The registers can be accessed via a simple register interface with address, data and read/write signals. The registers are 8 bits in width. To write to a register, \textit{rw\_in} must be set, \textit{addr\_in} should be driven by the target register address and \textit{data\_in} should hold the value of data that needs to be written to the register. To read from a register, \textit{rw\_in} must be de-asserted and the \textit{addr\_in} bus should be provided with the target register address. The \textit{data\_out} bus from the register interface will hold the value read from the target register. The register description is mentioned in Table \ref{adc_reg}. Register address 0 is a dummy register and kept for future use. Register address 1 is the configuration register where bit 0 is used to trigger an ADC conversion, bits 1 to 2 are used to set the oversampling factor, and bits 3 to 7 are reserved bits.

\subsection{ECC}

\acs{ECCs} are extensively used with the aim of protecting against soft-errors in automotive products that are crucial to safety. These errors arise either in the logic or in data due to radiation, electrical glitches, or electromagnetic interference, which can occur either during the production process or while the device is being used.

\begin{figure}[H]
	\centering
	\includegraphics[scale=0.65]{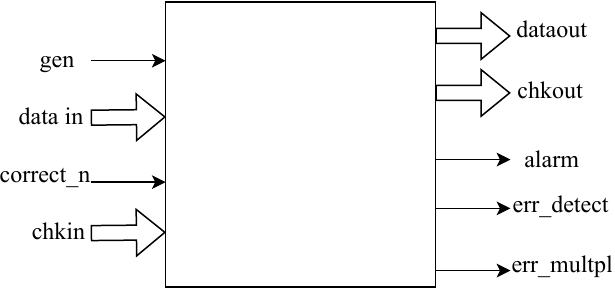}
	\caption{ECC Core}
	\label{design_ecc}
\end{figure}

The \ac{ECC} design implements a \ac{SECDED} over a parameterized range of word widths. It involves incorporating additional bits, called check bits, into each storage word (such as in RAM). During the writing process, these encoded check bits are written alongside the data bits. Upon reading a word, the check bits are utilized to determine if there are any errors and, if there is a single-bit error, which specific bit contains the error. This particular \ac{ECC} module is capable of detecting all one or two-bit errors (including check bits) and correcting all single-bit errors (including check bits). However, this \ac{ECC} does not guarantee error detection if more than two bits in a word are faulty (including check bits). In such cases, the error syndrome may match the error syndrome of a single-bit error, causing the \ac{ECC} to miscorrect a single-bit error that does not exist.

The block diagram of the \ac{ECC} core module is shown in Fig. \ref{design_ecc}. The same module can be configured as an encoder or a decoder based on the input signal \textit{gen}. When the \textit{gen} signal is set to 1, it configures the design as an encoder otherwise as a decoder. The bus \textit{datain} is the data word to check during the \textit{check mode}, or data from which check bits are generated during \textit{generate mode}. The primary output of the module is \textit{dataout} which can be corrected if an error is detected and \textit{correct\_n} is asserted. Other outputs include error check flags i.e., \textit{err\_detect} and \textit{err\_multpl} if the \ac{ECC} core module is used as decoder.

	\section{Verification Implementation}\label{sec:implementation}
\begin{figure}[H]
	\centering
	\includegraphics[scale=0.65]{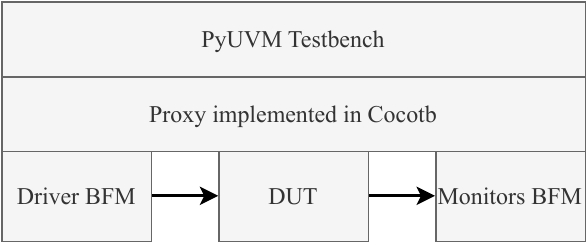}
	\caption{Proxy-driven PyUVM testbench \cite{Fitzpatrick} }
	\label{bfm_proxy}
\end{figure}

\begin{figure}[H]
	\centering
	\includegraphics[scale=0.65]{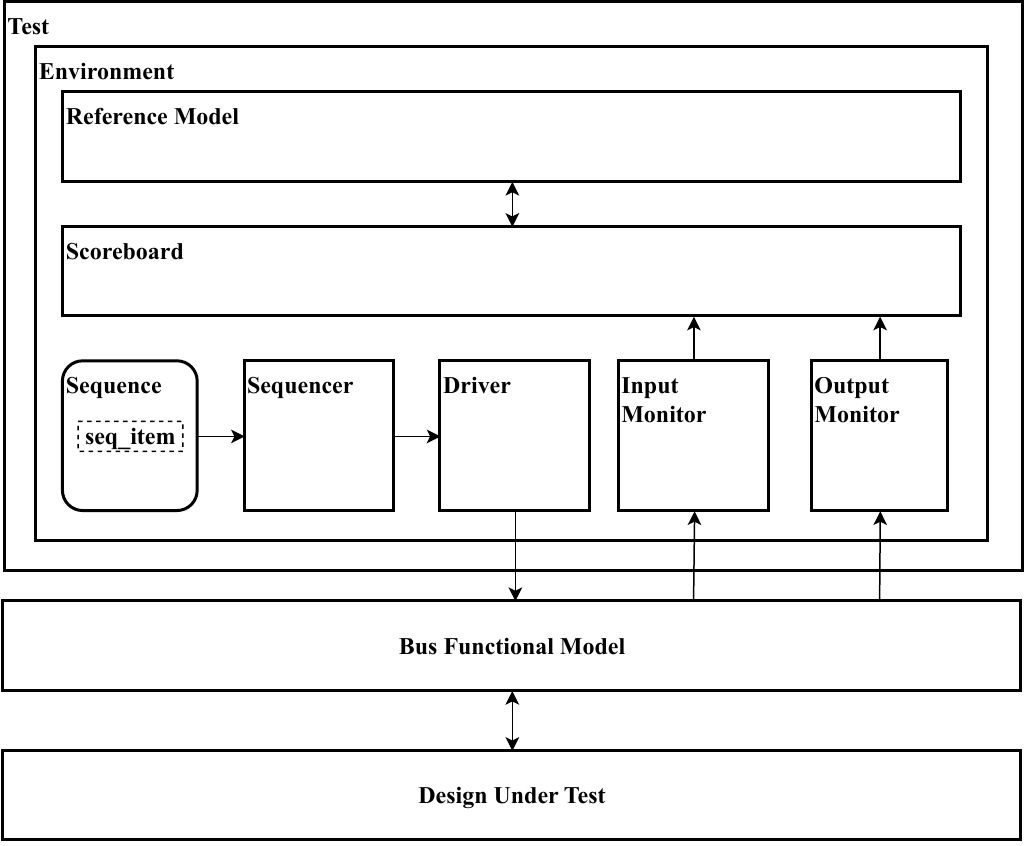}
	\caption{Testbench Architecture of PyUVM using \ac{BFM} class}
	\label{pyuvm_bd}
\end{figure}

	 With a clear understanding of the design \acp{IP}, we proceed with the verification implementation phase. In PyUVM as shown in Fig. \ref{bfm_proxy}, the testbench software is the PyUVM testbench, the proxy is implemented in Cocotb, where Cocotb connects Python to simulators through \ac{VPI} and \ac{VHPI}. Hence both Python and the \ac{DUT} which runs in the simulator share the same proxy interface, the transactions from the testbench are called using the Python coroutines (driver and monitors \acp{BFM}) which also ensures \ac{DUT} is not busy.
Figure \ref{pyuvm_bd} illustrates the PyUVM testbench architecture used in our work, which depicts the connections and communication between UVM components. The following subsections present a concise explanation of the PyUVM testbench implementation for the ECC design IP.

\subsection{ECC BFM}
To enable access to the \ac{DUT} and to abstract the Universal Verification Methodology (\ac{UVM}) testbench code, a \ac{BFM} class has been created. This \ac{BFM} has been implemented using coroutines in Cocotb. Within the \textit{\_\_init\_\_()} function of the \acp{BFM} class, a handle to the top module of the design is included. Additionally, a clock with a parameterized time period is generated as further explained in Listing \ref{initbfm}, line 2-3. To facilitate the transfer of transactions between the testbench and the \ac{DUT}, three queues have been declared, as specified in Listing \ref{initbfm}, line 4-6. 

\hspace{50pt}\begin{minipage}{.8\textwidth}
	\centering
	\begin{lstlisting}[language=Python, firstnumber=1, caption=\_\_init\_\_() function of ECC BFM class,label={initbfm},frame=tlrb]{Name}
def __init__(self):
	self.dut = cocotb.top
	cocotb.start_soon(Clock(self.dut.clk, CLKPERIOD, units="ns").start())
	self.driver_queue = Queue(maxsize=1)
	self.inp_mon_queue = Queue(maxsize=0) # Infinitely long
	self.out_mon_queue = Queue(maxsize=0) # Infinitely long
		
	\end{lstlisting}
\end{minipage}

As discussed before, the details of important coroutines inside the \ac{BFM}-based class are as below.

\subsubsection{initialization()}
The purpose of this function is to initialize the signals of the \ac{DUT}.

\subsubsection{driver\_bfm()} 
The implementation of the driver \ac{BFM} which is included in the \ac{BFM} class, is demonstrated in Listing \ref{driverbfm}. By utilizing the \textit{RisingEdge} and \textit{FallingEdge} triggers, this coroutine can function as an RTL \ac{BFM}. The \ac{BFM} is looped upon a clock positive edge. Within the \textit{try} block, a timed \textit{get()} operation is executed on a \textit{Queue} if a transaction can be found within it, whereas the loop will continue to the next iteration and wait for the next positive clock if there are no transactions in the \textit{Queue}. As shown in line 6, this transaction is assigned to the Encoder input signal \textit{datain}. In line 7 of Listing  \ref{driverbfm}, the \textit{gen\_random\_index(data\_in\_gen)} method is called which inserts bit flip(s) randomly in either data or check bits after encoding.

\hspace{50pt}\begin{minipage}{.8\textwidth}
	\centering
	\begin{lstlisting}[language=Python, firstnumber=1, caption= Driver coroutine in ECC BFM class,label={driverbfm},frame=tlrb]{Name}
async def driver_bfm(self):
	while True:
		await RisingEdge(self.dut.clk)
		try:
			self.data_in_gen = self.driver_queue.get_nowait()
			self.dut.enc_datain.value = self.data_in_gen
			self.gen_random_index(self.data_in_gen)
		except QueueEmpty:
			continue
		
	\end{lstlisting}
\end{minipage}
	
\subsubsection{inp\_mon\_bfm(), out\_mon\_bfm()} 
Two monitors are created utilizing coroutines from Cocotb in order to populate two infinite, long queues with input and output signals originating from the \ac{DUT}. The implementation details of these monitors can be found in Listing \ref{inmonitorbfm} and Listing \ref{outmonitorbfm}, respectively. These monitors are looped at the negative edge of the clock. The presence of queues permits the \ac{UVM} threads to make blocking calls into these asynchronous coroutines.

\hspace{50pt}\begin{minipage}{.8\textwidth}
	\centering
	\begin{lstlisting}[language=Python, firstnumber=1, caption=Input Monitor coroutine in ECC BFM class,label={inmonitorbfm},frame=tlrb]{Name}
async def inp_mon_bfm(self):
	while True:
		await FallingEdge(self.dut.clk)
		inp_tuple = (get_int(self.dut.enc_datain),get_int(self.dut.enc_chkin))
		self.inp_mon_queue.put_nowait(inp_tuple)
		
	\end{lstlisting}
\end{minipage}

\hspace{50pt}\begin{minipage}{.8\textwidth}
	\centering
	\begin{lstlisting}[language=Python, firstnumber=1, caption=Output Monitor coroutine in ECC BFM class,label={outmonitorbfm},frame=tlrb]{Name}
async def out_mon_bfm(self):
	while True:
		await FallingEdge(self.dut.clk)
		out_tuple = (get_int(self.dut.dec_err_detect), get_int(self.dut.dec_err_multpl), get_int(self.dut.dec_dataout), get_int(self.dut.dec_chkout))
		self.out_mon_queue.put_nowait(out_tuple)
		
	\end{lstlisting}
\end{minipage}

\subsubsection{start\_bfm()}
This method, which is also a part of the BFM class, is depicted in Listing \ref{startbfm} initiates the simultaneous execution of all coroutines during the run phase. It is launched in the build phase of the test, which is defined at the top of the testbench, as shown in Listing \ref{testpyuvm}, line 6.

\hspace{50pt}\begin{minipage}{.8\textwidth}
	\centering
	\begin{lstlisting}[language=Python, firstnumber=1, caption=Method to start all coroutines, label={startbfm},frame=tlrb]{Name}
def start_bfm(self):
	cocotb.start_soon(self.driver_bfm())
	cocotb.start_soon(self.inp_mon_bfm())
	cocotb.start_soon(self.out_mon_bfm())
		
	\end{lstlisting}
\end{minipage}

\subsection{Test}
Figure \ref{pyuvm_bd} depicts that the test comprises all the necessary UVM components essential for the testbench development. It is designated with the decorator\footnote{Decorators use @ followed by the decorator name before a function or class declaration. When invoking the decorated function or class, the decorator runs first, and its output substitutes the original.}, i.e., \textit{@pyuvm.test()}, to recognize itself as a test during the compilation process. The implementation of the test is shown in Listing \ref{testpyuvm}, which extends the \textit{uvm\_test}. 

During the build phase, the \ac{ECC} \ac{BFM} class object and \textit{EccEnv} are created, and the PyUVM version of ConfigDB, which is also a part of the \textit{uvm\_component}, is set.

The run phase in all UVM components is asynchronous and is declared as a coroutine\footnote{A coroutine function is a Python function declared with async def.} function using \textit{async def}. As it requires time, \textit{self.raise\_objection()} and \textit{self.drop\_objection()} are used to inform Cocotb of the simulation status (Started/Finished). Between these methods, a sequence is created and transferred to the sequencer to initiate, as shown in lines 12-14, Listing \ref{testpyuvm}.

\hspace{50pt}\begin{minipage}{.8\textwidth}
	\centering
	\begin{lstlisting}[language=Python, firstnumber=1, caption= Test top of PyUVM testbench,label={testpyuvm},frame=tlrb]{Name}
@pyuvm.test()
class BaseTestEcc(uvm_test):
	def build_phase(self):
		bfm = EccBfm()
		ConfigDB().set(None, "*", "BFM", bfm)        
		bfm.start_bfm()
		self.env = EccEnv.create("env", self)
	async def run_phase(self):
		self.raise_objection()
		seqr = ConfigDB().get(self, "", "SEQR")
		bfm = ConfigDB().get(self, "", "BFM")
		seq = EccSeq("seq")
		await seq.start(seqr)
		await ClockCycles(bfm.dut.clk, 3)
		self.drop_objection()
		
	\end{lstlisting}
\end{minipage}

\subsection{Environment}
The \textit{EccEnv} class is an extension of the \textit{uvm\_env} class, which encapsulates the necessary components required for the testbench, as outlined in Listing \ref{envpyuvm}. Within the build phase (Lines 3-8), an instance of the \textit{uvm\_sequencer} is created to queue and pass sequence items to the driver, and the Driver, Monitors, and Scoreboard are instantiated. In the Monitors, the name of the proxy functions that get the data they monitor is passed. 

During the connect phase, the exports are connected to the ports, as exemplified in lines 10-12, Listing \ref{envpyuvm}. The build phase is a top-down phase, while the connect phase is a bottom-up phase.

\hspace{50pt}\begin{minipage}{.8\textwidth}
	\centering
	\begin{lstlisting}[language=Python, firstnumber=1, caption=Environment of PyUVM testbench,label={envpyuvm},frame=tlrb]{Name}
class EccEnv(uvm_env):
	def build_phase(self):
		self.seqr = uvm_sequencer("seqr", self)
		ConfigDB().set(None, "*", "SEQR", self.seqr)
		self.driver = Driver("driver", self)
		self.inp_mon = InputMonitor("inp_mon", self, "get_inp")
		self.out_mon = OutputMonitor("out_mon", self, "get_out")
		self.scoreboard = Scoreboard("scoreboard", self)
	def connect_phase(self):
		self.driver.seq_item_port.connect(self.seqr.seq_item_export)
		self.inp_mon.ap.connect(self.scoreboard.inp_export)
		self.out_mon.ap.connect(self.scoreboard.out_export)

	\end{lstlisting}
\end{minipage}

\subsection{Monitor}
\label{monitor}
Figure \ref{pyuvm_bd} demonstrates that the PyUVM testbench implementation may include two monitors, namely the \textit{Input Monitor} and \textit{Output Monitor}, which can monitor input and output transactions from the \ac{DUT} separately. Listing \ref{monpyuvm} illustrates the code for the \textit{OutputMonitor} class, which is an extension of the \textit{uvm\_component}. It accepts the name of the Cocotb proxy method as an argument, as presented in Line 2. 

During the build phase, it utilizes this argument to locate the method in the proxy and subsequently invokes the method utilizing \textit{getattr()}, as demonstrated in line 8. This functionality is infeasible in SystemVerilog-\ac{UVM}. Additionally, an analysis port is created. Within the run phase, an instance of a covergroup is generated, which is sampled with the datum at each clock edge. Moreover, data is written into the analysis port, which is obtained at each clock edge, as displayed in line 14.

\hspace{50pt}\begin{minipage}{.8\textwidth}
	\centering
	\begin{lstlisting}[language=Python, firstnumber=1, caption= Monitor of PyUVM testbench,label={monpyuvm},frame=tlrb]{Name}
class OutputMonitor(uvm_component):
	def __init__(self, name, parent, method_name):
		super().__init__(name, parent)
		self.method_name = method_name
	def build_phase(self):
		self.ap = uvm_analysis_port("ap", self)
		self.bfm = ConfigDB().get(self, "", "BFM")
		self.get_method = getattr(self.bfm, self.method_name)
	async def run_phase(self):
		dut_cg = dut_covergroup()
		while True:
			datum = await self.get_method()
			dut_cg.sample(datum)
			self.ap.write(datum)

	\end{lstlisting}
\end{minipage}

\subsection{Scoreboard}
The \textit{Scoreboard} class, detailed in Listing \ref{scrbdpyuvm}, also extends the \textit{uvm\_component} class, similar to monitors. In essence, this class receives inputs and outputs from the \ac{DUT}; and analyzes and verifies the proper functioning of the \ac{DUT}.

During the build phase (Lines 3-8), instances of the \textit{uvm\_tlm\_analysis\_fifo} are created to obtain data from monitors and store them. In the connect phase, the exports are connected, as demonstrated in lines 10 and 11. After the simulation run, the check phase is executed. Using the \textit{try\_get()} method, which returns a tuple consisting of data retrieval success (True/False) and the data, data is retrieved from both input and output FIFOs that are iterated conditionally using \textit{can\_get()}. Within the checker function as demonstrated in lines 18-22, a comparison is made to determine whether the encoder data input matches the decoder data output. Otherwise, there might be multiple bit flips (as the \ac{ECC} is \ac{SECDED}).

At the conclusion of the simulation, the coverage report is generated utilizing the \textit{write\_coverage\_db} method from PyVSC, as shown in line 23, Listing \ref{scrbdpyuvm}.

\hspace{50pt}\begin{minipage}{.8\textwidth}
	\centering
	\begin{lstlisting}[language=Python, firstnumber=1, caption=Scoreboard of PyUVM Testbench,label={scrbdpyuvm},frame=tlrb]{Name}
class Scoreboard(uvm_component):
	def build_phase(self):
		self.inp_fifo = uvm_tlm_analysis_fifo("inp_fifo", self)
		self.out_fifo = uvm_tlm_analysis_fifo("out_fifo", self)
		self.inp_get_port = uvm_get_port("inp_get_port", self)
		self.out_get_port = uvm_get_port("out_get_port", self)
		self.inp_export = self.inp_fifo.analysis_export
		self.out_export = self.out_fifo.analysis_export
	def connect_phase(self):
		self.inp_get_port.connect(self.inp_fifo.get_export)
		self.out_get_port.connect(self.out_fifo.get_export)
	def check_phase(self):
		while self.out_get_port.can_get():
			_, dut_out = self.out_get_port.try_get()
			inp_success, inp = self.inp_get_port.try_get()
			(enc_datain, enc_chkin) = inp
			(dec_err_det, dec_err_multpl, dec_dataout, dec_chkout, dec_alarmout) = dut_out
			if inp_success == True:
				if dec_dataout == enc_datain:
					self.logger.info(f"Test: Yay!!! Passed!")
				else:
					self.logger.debug(f"Test: Failed! Decoded data mismatched!! Multiple bit flips exist!!! Expected: {enc_datain}, Actual: {dec_dataout}")
		vsc.write_coverage_db('cov.xml')

	\end{lstlisting}
\end{minipage}

\subsection{Driver}
Listing \ref{drvpyuvm} illustrates the implementation of the PyUVM driver component. The \textit{Driver} class extends the \textit{uvm\_driver} and works with sequence items. The \ac{ECC} \ac{BFM} method is accessed using ConfigDB. Therefore, the \textit{get()} method is implemented in the connect phase, as shown in line 3.

In the run phase (Lines 5-9, Listing \ref{drvpyuvm}), the initialization function from the \ac{ECC} \ac{BFM} class is invoked, and then the \textit{get\_next\_item()} method, which is defined inside an infinite loop, is utilized to retrieve the sequence items and transmit them to the \textit{driver\_bfm} function in the \ac{BFM} class by calling \textit{send\_inp}.

\hspace{50pt}\begin{minipage}{.8\textwidth}
	\centering
	\begin{lstlisting}[language=Python, firstnumber=1, caption=Driver of PyUVM testbench,label={drvpyuvm},frame=tlrb]{Name}
class Driver(uvm_driver):
	def connect_phase(self):
		self.bfm = ConfigDB().get(self, "", "BFM")
	async def run_phase(self):
		self.bfm.initialization()
		while True:
			input = await self.seq_item_port.get_next_item()
			await self.bfm.send_inp(input.datain)
			self.seq_item_port.item_done()

	\end{lstlisting}
\end{minipage}

\subsection{Sequence}
The \textit{EccSeq} class, detailed in Listing \ref{seqpyuvm}, extends the \textit{uvm\_sequence} class. It contains a body that generates sequence items, randomizes them, and transmits them to the Driver. Since the \textit{await} keyword is utilized, the \textit{start\_item} and \textit{finish\_item} methods wait for access to the sequencer and transmit the items to the driver, respectively. The number of transactions can be specified in the loop count, as demonstrated in line 3.

\hspace{50pt}\begin{minipage}{.8\textwidth}
	\centering
	\begin{lstlisting}[language=Python, firstnumber=1, caption=A sequence in PyUVM testbench ,label={seqpyuvm},frame=tlrb]{Name}
class EccSeq(uvm_sequence):
	async def body(self):
		for i in range(30000):
			in_tr = EccSeqItem("in_tr")
			await self.start_item(in_tr)
			in_tr.randomize()
			await self.finish_item(in_tr)
	\end{lstlisting}
\end{minipage}

\subsection{Sequence Item}
Listing \ref{seqitempyuvm} shows code for \textit{EccSeqItem} which extends \textit{uvm\_sequence\_item}. In the listing, lines 5-7 show that it includes all the stimuli declaration using PyVSC as specified by \textit{@vsc.randobj} decorator at the top of class definition. 

The \textit{\_\_eq\_\_()}, \textit{\_\_str\_\_()} methods are defined to compare and print string version of items respectively as shown in lines 9-13, Listing \ref{seqitempyuvm}.

\hspace{50pt}\begin{minipage}{.8\textwidth}
	\centering
	\begin{lstlisting}[language=Python, firstnumber=1, caption=Sequence item defined in PyUVM testbench,label={seqitempyuvm},frame=tlrb]{Name}
@vsc.randobj
class EccSeqItem(uvm_sequence_item):
	def __init__(self, name):
		super().__init__(name)
		self.correct_n = vsc.bit_t(1)
		self.datain = vsc.rand_uint32_t()
		self.chkin = vsc.bit_t(7)
		...
	def __eq__(self, other):
		same = self.correct_n == other.correct_n and self.datain == other.datain and ...
		return same
	def __str__(self):
		return f"{self.get_name()} : correct_n: {self.correct_n}}"
	\end{lstlisting}
\end{minipage}


	\section{Results}\label{sec:results}
	The verification testbenches for the aforementioned  design \ac{IPs} are implemented using SystemVerilog-\ac{UVM} and PyUVM. Specifically, PyVSC library is used along with PyUVM which enables constrained randomization and functional coverage constructs \cite{Ballance} for Python testbenches. The results produced using both methodologies are analyzed, compared with respect to simulation performance and features. 

\subsection{Feature Comparison}
In our work, certain features are compared between SystemVerilog-\ac{UVM} and PyUVM implementations as shown in Table \ref{comp_feature}. Additionally, it is found that the design hierarchy for the PyUVM testbench does not include the top testbench in the simulator tools which somehow hinders its debugging capabilities. But SystemVerilog-\ac{UVM} includes the top module, \textit{uvm\_test\_top} along with its inner components in the simulation or debugging tools. The detailed explanation is also discussed in the work \cite{suruchi}.

\begin{table}[H]
	
	\centering
	\caption{Comparison between SystemVerilog-UVM and PyUVM}
	\label{comp_feature}
	\resizebox{\linewidth}{!}{%
		\begin{tabular}{|>{\hspace{0pt}}p{0.154\linewidth}|>{\hspace{0pt}}p{0.181\linewidth}|>{\hspace{0pt}}p{0.188\linewidth}|>{\hspace{0pt}}p{0.469\linewidth}|} 
			\hline
			\textbf{Feature} & \textbf{SystemVerilog-UVM} & \textbf{PyUVM} & \textbf{Remarks} \\ 
			\hline
			\textbf{Utility Macros} & \`{}uvm\_object\_utils,\par{}~\`{}uvm\_object\_utils\_begin, \par{}\`{}uvm\_object\_utils\_end~ & - & In SystemVerilog-UVM, macros are used to register \par{}classes in the UVM factory, whereas in PyUVM, \par{}All classes are already registered. \\ 
			\hline
			\textbf{Field Macros} & ~\`{}uvm\_field\_*~ & - & In SystemVerilog-UVM, macros implement methods like~ \par{}do\_compare, convert2string(). In PyUVM, \_\_str\_\_() and \par{}\_\_eq\_\_() are used for converting to string and comparing\par{}~respectively. \\ 
			\hline
			\textbf{UVM RAL Model} & Provides standard base \par{}class libraries & - & In
			PyUVM, the implementation of RAL model is in pipeline. \\ 
			\hline
			\textbf{Logging} & Implements UVM \par{}Reporting System & Uses
			logging module & In PyUVM, reporting method adds two extra logging levels: \par{}FIFO\_DEBUG (5) and~ PYUVM\_DEBUG (4) \\ 
			\hline
			\textbf{ConfigDB} & uvm\_config\_db\_\#int :: set\par{}(null, "*", 'BFM',bfm)~ & ConfigDB().set(None,\par{}~"*","BFM", bfm) & Instead of static function calls and long incantation with various types, \par{}PyUVM provides a singleton that stores data in the configuration \par{}database using the same hierarchy-based control as the \par{}SystemVerilog version. Also, PyUVM eases debugging of ConfigDB(). \\ 
			\hline
			\textbf{Importation} & import
			uvm\_pkg::* & from
			pyuvm import * & All the UVM classes and functions become available in our code \par{}without referencing the package. \\ 
			\hline
			\textbf{Instantiating }\par{}\textbf{objects using }\par{}\textbf{the factory} & env = alu\_env::type\_id::\par{}create("env", this) & self.env = AluEnv.create\par{}("env",self) & In
			PyUVM, there is no type involved as it directly copies the handles. \\ 
			\hline
			\textbf{uvm\_subscriber }\par{}\textbf{class} & class uvm\_subscriber \par{}\#(type T =int) extends \par{}uvm\_component & class Subscriber\par{}(uvm\_analysis\_export) & In SystemVerilog-UVM, uvm\_subscriber creates an analysis\_export \par{}with the correct parameterized type. In PyUVM, since Python does not\par{}have typing issues, a subscriber can be created by directly extending \par{}uvm\_analysis\_export and providing the write() function. \\ 
			\hline
			\textbf{User-defined phases} & Possible & Not
			possible & PyUVM
			only implements common phases of UVM specification. \\ 
			\hline
			\textbf{uvm\_test} & Required & Not
			required & PyUVM does not need uvm\_test, though it implements it to follow the \par{}standard with an empty extension. In run\_test, class can be passed directly. \\ 
			\hline
			\textbf{Awaiting tasks} & - & await
			run\_test() & In SystemVerilog-UVM, there is no indication for time-consuming calls, \par{}whereas in PyUVM, such calls are done using await. \\ 
			\hline
			\textbf{TLM System} & *\_imp
			classes & - & In SystemVerilog-UVM, the *\_imp classes are required to provide \par{}implementations of tasks such as put() and get(). In PyUVM, any \par{}uvm\_component can instantiate uvm\_put\_port or uvm\_get\_port in \par{}its build\_phase(). \\
			\hline
		\end{tabular}
	}

\end{table}

\subsection{Performance Metrics}
\subsubsection{Simulation Runtime}
\begin{figure}[H]
	\centering
	\begin{subfigure}{0.495\textwidth}
		\includegraphics[width=1\linewidth]{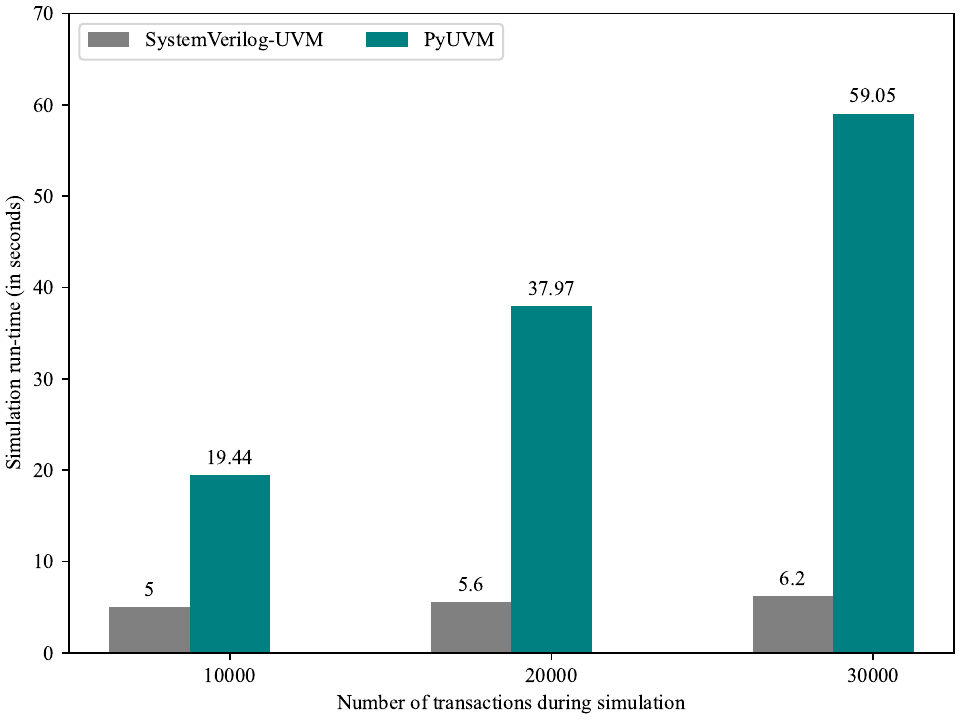} 
		\caption{ALU: A test is run separately for various transactions}
		\label{run_alu}
	\end{subfigure}
	\hfill
	\begin{subfigure}{0.495\textwidth}
		\includegraphics[width=1\linewidth]{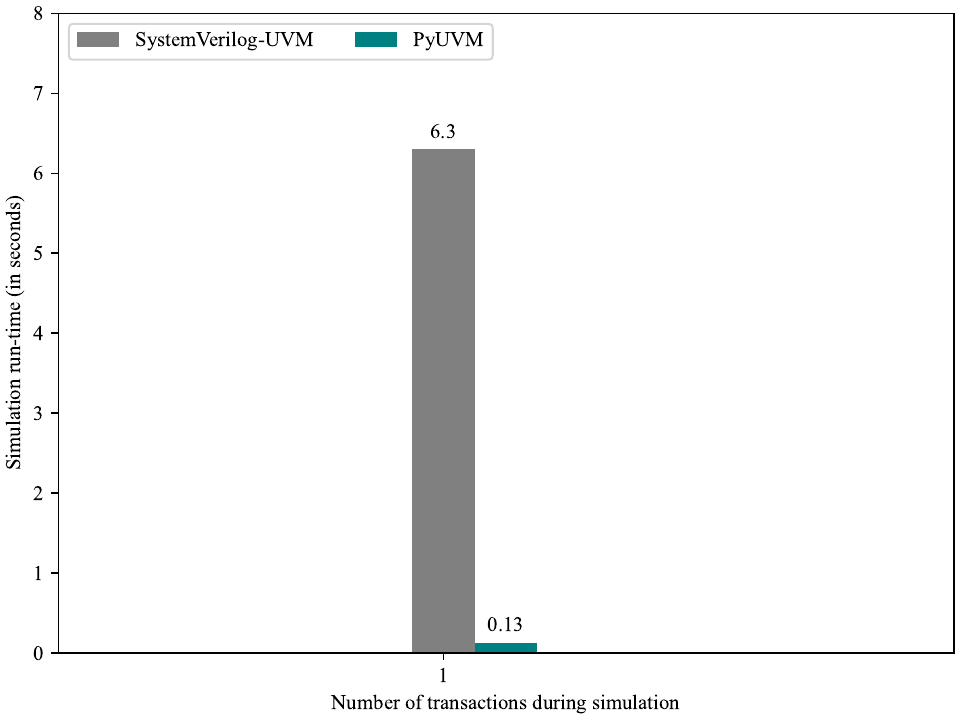}
		\caption{ADC: 3 tests are run in a single transaction}
		\label{run_i2c}
	\end{subfigure}
	\begin{subfigure}{0.495\textwidth}
		\centering
		\vspace{10pt}
		\includegraphics[width=1\linewidth]{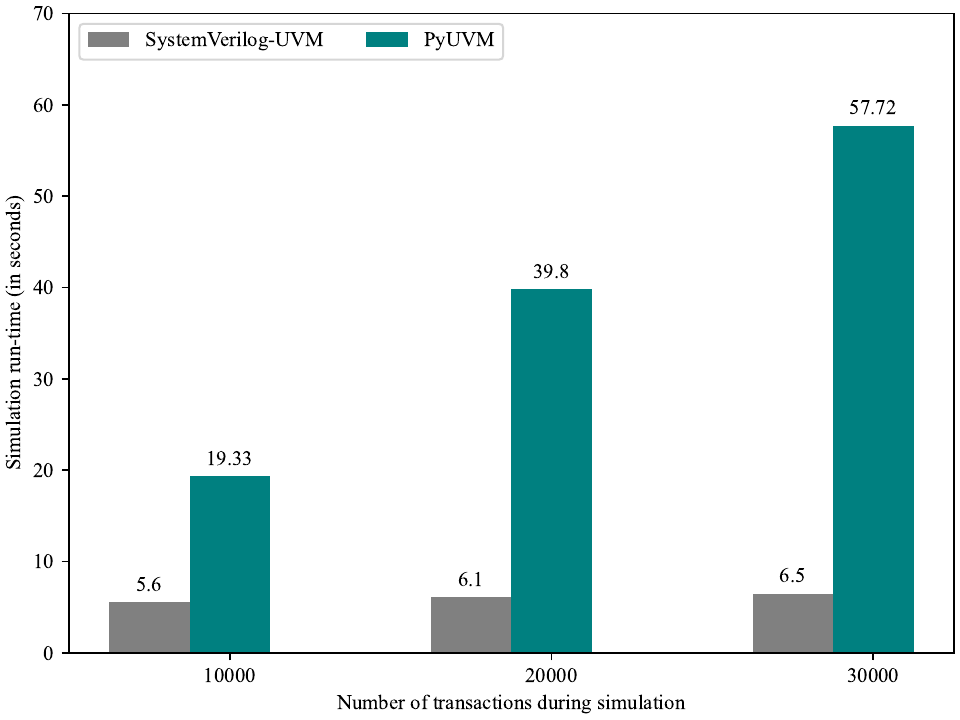}
		\caption{ECC: A test is run separately for various transactions}
		\label{run_adc}
	\end{subfigure}
	
	\caption{Simulation run-time between PyUVM and SystemVerilog-UVM}
	\label{run_time}
\end{figure}

The PyUVM and SystemVerilog-\ac{UVM} testbenches are compared in terms of simulation run-time with simulator Cadence Xcelium. For both \ac{ALU} and \ac{ECC}, a single test specified in the testbenches is simulated with various iterations i.e., 10000, 20000, and 30000 as shown in Fig.\ref{run_alu} and Fig. \ref{run_adc}. ADC testbenches include 3 tests, namely \textit{test\_feature\_adc}, \textit{test\_feature\_reg}, and \textit{test\_stress\_adc}. It is also ensured to keep all the simulation set-up parameters constant for both PyUVM and SystemVerilog-\ac{UVM} testbenches to make a fair comparison.

Fig.\ref{run_alu} and Fig.\ref{run_adc} show that the simulation run-time of PyUVM testbenches is slower than that of SystemVerilog-\ac{UVM} testbenches for all iterations. This contrast in run-time performance between SystemVerilog-UVM and Python testbenches is due to their distinct approaches. With SystemVerilog, simulation directives and commands are employed to establish communication with the simulator, resulting in a close integration that enhances execution and shortens run-time. In contrast, Python testbenches interact with the simulator using VPI/VHPI, which is slower and less tightly integrated. This overhead becomes more significant as the number of transactions increases, leading to longer simulation run-time for PyUVM testbenches.

In contrast, PyUVM performs better than SystemVerilog-\ac{UVM} in terms of simulation run-time for \ac{ADC} as demonstrated in Fig.\ref{run_adc}. PyUVM which uses Cocotb, automatically discovers all tests defined with the help of the \textit{@pyuvm.test()} decorator. Therefore it may help speeding up the simulation which is not the case in SystemVerilog-\ac{UVM}. 

\subsubsection{Coverage Analysis}
To have a fair comparison between PyUVM and SystemVerilog-\ac{UVM}, testbenches for previously mentioned \ac{IPs} are implemented in SystemVerilog-\ac{UVM} with the same coverage model. The model is defined as such that it covers all the functionalities of the design \ac{IPs} as discussed in Table \ref{cov_mod}.

\begin{table}
	\centering
	\caption{Coverage model definition for the designs}
	\setlength{\extrarowheight}{3pt}
	\setlength{\tabcolsep}{5pt}
	\label{cov_mod}
	\begin{tabular}{|c|c|l|l|l|} 
		\hline
		\multicolumn{1}{|l|}{\begin{tabular}[c]{@{}l@{}}\textbf{Design }\\\textbf{IP}\end{tabular}} & \multicolumn{1}{l|}{\begin{tabular}[c]{@{}l@{}}\textbf{Cover}\\\textbf{Group(s)}\end{tabular}} & \textbf{Cover Point(s)} & \begin{tabular}[c]{@{}l@{}}\textbf{Number of }\\\textbf{Bins}\end{tabular} & \textbf{Description of Cover Bin(s)}                                                                                                                                                                                                           \\ 
		\hline
		\multirow{7}{*}{\textbf{ALU}}                                                               & \multirow{7}{*}{cg\_1~}                                                                        & a                       & \multirow{2}{*}{5~}                                                        & \begin{tabular}[c]{@{}l@{}}Value ranges (-2147483648, -1768769053), \\(-1768769052, -866), (-865, 866), \\(867, 1300000000), (1300000001, 2147483647)\end{tabular}                                                                             \\ 
		\cline{3-3}\cline{5-5}
		&                                                                                                & b                       &                                                                            & \begin{tabular}[c]{@{}l@{}}Value ranges (-2147483648, -1654895901), \\(-1654895902, -989), (-988, 0), \\(1, 1928710300), (1928710301, 2147483647)\end{tabular}                                                                                 \\ 
		\cline{3-5}
		&                                                                                                & op                      & 8~                                                                         & Cover all 8 operations                                                                                                                                                                                                                          \\ 
		\cline{3-5}
		&                                                                                                & aXb                     & \multirow{4}{*}{-}                                                         & \multirow{4}{*}{-}                                                                                                                                                                                                                             \\ 
		\cline{3-3}
		&                                                                                                & bXop                    &                                                                            &                                                                                                                                                                                                                                                \\ 
		\cline{3-3}
		&                                                                                                & aXop                    &                                                                            &                                                                                                                                                                                                                                                \\ 
		\cline{3-3}
		&                                                                                                & aXbXop                  &                                                                            &                                                                                                                                                                                                                                                \\ 
		\hline
		\multirow{9}{*}{\textbf{ADC}}                                                               & \multirow{5}{*}{cg\_1~}                                                                        & en\_in                  & 2                                                                          & ADC module is ON (1),~ OFF (0)                                                                                                                                                                                                                 \\ 
		\cline{3-5}
		&                                                                                                & rw\_in                  & 2                                                                          & Register Write (1), Read (0)                                                                                                                                                                                                                   \\ 
		\cline{3-5}
		&                                                                                                & addr\_in                & \multirow{3}{*}{5~}                                                        & \multirow{2}{*}{Value ranges from 0 to 255}                                                                                                                                                                                                    \\ 
		\cline{3-3}
		&                                                                                                & data\_in                &                                                                            &                                                                                                                                                                                                                                                \\ 
		\cline{3-3}\cline{5-5}
		&                                                                                                & ana\_in                 &                                                                            & Scope: -10V to10V                                                                                                                                                                                                                              \\ 
		\cline{2-5}
		& \multirow{4}{*}{cg\_2}                                                                         & start\_out              & \multirow{4}{*}{2~}                                                        & Conversion is started or not                                                                                                                                                                                                                    \\ 
		\cline{3-3}\cline{5-5}
		&                                                                                                & busy\_out               &                                                                            & Transaction is ongoing                                                                                                                                                                                                                         \\ 
		\cline{3-3}\cline{5-5}
		&                                                                                                & eoc\_out                &                                                                            & Conversion is ended or not                                                                                                                                                                                                                     \\ 
		\cline{3-3}\cline{5-5}
		&                                                                                                & err\_out                &                                                                            & Covers if there is errored conversion                                                                                                                                                                                                           \\ 
		\hline
		\multirow{5}{*}{\textbf{ECC}}                                                               & \multirow{5}{*}{cg\_1~}                                                                        & data\_out               & 8~                                                                         & \begin{tabular}[c]{@{}l@{}}Value ranges (0, 2975706), (2975707, 10295960), \\(10295961, 56784980), (56784981, 130000000), \\(130000001, 78939421), (789394220, 1248579698), \\(1248579699, 2000000000), (2000000001, 2147483647)\end{tabular}  \\ 
		\cline{3-5}
		&                                                                                                & chkout                  & 3~                                                                         & Cover ranges (0, 23),(24, 89), (90, 127)                                                                                                                                                                                                       \\ 
		\cline{3-5}
		&                                                                                                & err\_detect             & \multirow{2}{*}{2~}                                                        & Error is detected                                                                                                                                                                                                                              \\ 
		\cline{3-3}\cline{5-5}
		&                                                                                                & err\_multpl             &                                                                            & Multiple error exists                                                                                                                                                                                                                          \\ 
		\cline{3-5}
		&                                                                                                & err\_detectXerr\_multpl & -                                                                          & -                                                                                                                                                                                                                                              \\
		\hline
	\end{tabular}
\end{table}

Our study demonstrates the feasibility of creating coverage models in PyUVM testbenches using the PyVSC library in comparison to SystemVerilog-\ac{UVM}. However, there is no evidence to suggest that PyUVM outperforms SystemVerilog-\ac{UVM} in terms of coverage closure. Nonetheless, the PyVSC library could be employed to restrict the stimuli and close the coverage of the \ac{DUT}. The results of our simulations indicate that a \SI{100}{\percent} coverage in PyUVM simulations of the \ac{ALU} design \ac{IP} may have resulted from the random seed selected, as demonstrated in Table \ref{comp_coverage}.

\begin{table}[H]
	\centering
	\setlength{\extrarowheight}{3pt}
	\setlength{\tabcolsep}{5pt}
	\caption{Coverage analysis between SystemVerilog-UVM and PyUVM}
	\label{comp_coverage}
	\begin{tabular}{|l|rr|cc|rr|}
		\hline
		\textbf{Design IP}                & \multicolumn{2}{c|}{\textbf{ALU}}                                          & \multicolumn{2}{c|}{\textbf{ADC}}                               & \multicolumn{2}{c|}{\textbf{ECC}}                                          \\ \hline
		\textbf{SV-UVM/ PyUVM}            & \multicolumn{1}{c|}{\textbf{SV-UVM}} & \multicolumn{1}{c|}{\textbf{PyUVM}} & \multicolumn{1}{c|}{\textbf{SV-UVM}} & \textbf{PyUVM}           & \multicolumn{1}{c|}{\textbf{SV-UVM}} & \multicolumn{1}{c|}{\textbf{PyUVM}} \\ \hline
		\textbf{Number of Distinct Tests} & \multicolumn{1}{r|}{1}               & 1                                   & \multicolumn{1}{r|}{3}               & \multicolumn{1}{r|}{3}   & \multicolumn{1}{r|}{1}               & 1                                   \\ \hline
		\textbf{Number of Transactions}   & \multicolumn{1}{r|}{30000}           & 30000                               & \multicolumn{1}{r|}{-}               & \multicolumn{1}{r|}{-}                        & \multicolumn{1}{r|}{30000}           & 30000                               \\ \hline
		\textbf{Coverage (\%)}            & \multicolumn{1}{r|}{78.29}           & 100                                 & \multicolumn{1}{r|}{100}               & \multicolumn{1}{r|}{100} & \multicolumn{1}{r|}{95}              & 95                                  \\ \hline
	\end{tabular}
\end{table}

\subsection{Empirical Observations}
While implementing the verification enivironment for the design IPs i.e., ALU, ADC, and ECC with PyUVM, there are some observations made as listed below.

\subsubsection{BFM-based class approach}
In SystemVerilog-\ac{UVM}, an interface is used to send transactions/packets to the \ac{DUT}. The declaration of stimuli inside the interface is necessary to accomplish this. On the other hand, PyUVM uses a \ac{BFM}-based approach for communication between the \ac{DUT} and testbench, which utilizes Cocotb. Therefore, no additional stimuli declaration is required.

\subsubsection{Register Abstraction Layer Modeling}
In SystemVerilog-\ac{UVM}, the \textit{UVM-RAL} models and abstracts registers and memories of a \ac{DUT}. In our work, one of the design \ac{IP} i.e., \ac{ADC}, includes a register block with 3 registers as explained in Table \ref{adc_reg}. Because of unavailability of the \textit{UVM-RAL}, constrained randomization in the PyUVM testbench requires additional efforts to read and write to these registers, whereas in SystemVerilog-\ac{UVM}, this process is simplified through the use of \textit{UVM-RAL}. It should be noted, however, that RAL is mentioned as "under development" as listed in Table \ref{comp_feature}.

\subsubsection{Ease of Testbench Development}
As explained in the feature comparison, in contrast to SystemVerilog-\ac{UVM}, PyUVM does not need a class constructor. Additionally, SystemVerilog does not allow introspection whereas PyUVM enables it for better coding style. For instance, the Monitor takes a proxy method name as an argument during instantiation. In the run phase of the monitor, the data can be transferred using \textit{get\_method()}, detailed in subsection \ref{monitor}. Overall, PyUVM needs less code lines compared to SystemVerilog-\ac{UVM}. 

\subsubsection{Continuous Assignment}
The testbench implementation of ECC core needs to continuously assign encoded data from the encoder to the decoder as input. Using the \textit{assign} keyword in SystemVerilog-\ac{UVM}, the assignment can be accomplished as shown in Listing \ref{contasssv}. On the other hand, during the PyUVM testbench implementation of \ac{ECC}, it is found that \textit{data\_out} is not assigned correctly with the command as mentioned in the listing \ref{contasspy}. Consequently, instances of Encoder and Decoder are separated and encoded data is sent from the testbench instead of continuously assigning with respect to each clock edge.

\hspace{50pt}\begin{minipage}{.8\textwidth}
	\centering
	\begin{lstlisting}[language=Python, numbers=none, caption=Continuous assignment in SystemVerilog,label={contasssv},frame=tlrb]{Name}
assign data_input = {vif_encoder.dataout, vif_encoder.chkout};
	\end{lstlisting}
\end{minipage}

\hspace{50pt}\begin{minipage}{.8\textwidth}
	\centering
	\begin{lstlisting}[language=Python,  numbers=none, caption=Continuous assignment in Python,label={contasspy},frame=tlrb]{Name}
data_input = self.dut.enc_dataout.value.binstr + self.dut.enc_chkout.value.binstr
	\end{lstlisting}
\end{minipage}

\subsubsection{PyVSC usage}
The coverage construct is already available in SystemVerilog. On the other hand, Python as \ac{HVL} does not have a covergroup. Hence, PyVSC is used for constrained randomization and writing coverage constructs. It allows saving the coverage results in a \textit{.xml} or \textit{.libucis} format, which can be visualized using \textit{PyUCIS-Viewer} \cite{Pyucis}.

	\section{Conclusion}\label{sec:conclusion}
	In this study, we have developed Python-based verification testbenches for three design IPs: \ac{ALU}, \ac{ADC}, and \ac{ECC}. The testbenches have been implemented by employing the PyUVM framework and PyVSC library. The comparison between PyUVM and SystemVerilog-\ac{UVM} has been carried out in terms of various features. Furthermore, the performance of PyUVM testbenches has been evaluated and compared with SystemVerilog-\ac{UVM} testbenches for the aforementioned designs. Despite taking longer to run, PyUVM simulation may be more efficient as compared to SystemVerilog-\ac{UVM}, provided that the clock generation is moved from the testbench to the \ac{DUT} side. PyUVM simulation enabled us to conveniently collect input data along with coverage data in a preferred format. This data could be analyzed to create new methodologies based on Machine Learning techniques, which will further enhance the design verification process.

	\label{sec:bibliography}
	\printbibliography[heading=bibintoc]%

\end{document}